\documentclass[doublecol]{epl2} 
\usepackage{graphicx,makeidx,color}
\usepackage{amsmath}                                  
\usepackage{epsfig}
\usepackage{amssymb} 
\usepackage{color}
\title{Evolution  and    non-equilibrium  physics. A study of the Tangled Nature Model}
\author{Nikolaj Becker and Paolo Sibani }

\institute{                    
   FKF,  University of Southern Denmark, Campusvej 55, DK5230, Odense M.
  }
\pacs{87.23.Kg}{Evolution in biology}
\pacs{89.75-k}{Complex systems}
\pacs{89.75.Fb}{Structure and organization in complex systems}

\date{\today}

\abstract{
We argue that the  stochastic dynamics of interacting agents which replicate, mutate and die
constitutes a  non-equilibrium  physical process akin to aging in complex materials.
Specifically, our study uses  extensive computer  simulations  of the Tangled Nature Model (TNM)
of biological evolution to show    that    punctuated
equilibria successively generated  by   the  model's dynamics  have increasing entropy and are separated by
increasing entropic barriers. We further  show that these states are organized in a hierarchy  { and
that limiting the values of possible interactions to a finite interval   leads to stationary 
fluctuations within  a component of the latter}.
A coarse-grained     description based
on the  temporal statistics of  quakes, the events leading from one component of the hierarchy to the next,
accounts for the logarithmic growth of the population and the decaying rate of 
change of macroscopic variables.
Finally, we question  the role of fitness in  large scale evolution models  and speculate on the possible evolutionary role of  rejuvenation and memory effects. 
}  
\begin{document}
 \maketitle

\noindent {\bf Introduction.} 
Initially perceived as a challenge to gradualism, punctuated equilibria
 are now widely accepted~\cite{Gould77,Gould02} as  key features of large scale darwinian evolution. 
Their striking  similarity  to  intermittency in   `aging'~\cite{Sibani89,Oliveira05,Sibani06a,Boettcher11}
complex materials  is not well understood, but may hold clues   on how life  evolves from  matter~\cite{Trefil09}. 
The  origin of this similarity  is addressed  below  by analyzing  the Tangled Nature Model (TNM)
dynamics~\cite{Christensen02,Anderson05}
 as  a   non-equilibrium physical process.

 While   physics ideas are common in evolution  models~\cite{Black12,Drossel01},  evolution itself 
 has  not  previously  been modeled   as a physical process,  bar 
 attempts~\cite{Bak93,Bak97,Sneppen95}  inspired by Self Organized Criticality (SOC)~\cite{Bak87},
  according to which  punctuations are the manifestation of stationary  fluctuations.
We   see  them instead  as  the manifestation of 
 a  spontaneous   physical process.
But how can   a pertinent   free energy then be defined and why does the
process   decelerate   over time~\cite{Newman99e,Alroy08}?

In spite of its simplicity, the TNM, an individual based stochastic model
of ecosystem dynamics,  captures key aspects of co-evolution, e.g. its decelerating
 nature~\cite{Jones10}, its log-normal species abundance distribution~\cite{Hall02}
and,  in a version including spatial migration, the area law~\cite{Lawson06}. Punctuations, here called  \emph{quakes}, 
irreversibly disrupt \emph{quasi-Evolutionary Stable Strategies} (qESS), 
periods of metastability where   population and  the number of extant species, or diversity, fluctuate  reversibly. 
Statistical physics is used to connect  microscopic interactions, defined in darwinian terms at the level of individuals, to   macroscopic properties,
e.g. population and diversity. Along the way, we introduce the concepts of \emph{core} and \emph{cloud} species and implement 
an adaptation of the \emph{lid method}~\cite{Sibani99} originally developed to map out complex energy landscapes. We find
 that: \emph{i)} The growing duration  of qESS reflects an entrenchment into metastable configuration space
 components of  increasing entropy; 
\emph{ii)} The decreasing rate of   evolution stems from a logarithmic time growth of 
the entropic barriers separating successive qESS;
\emph{iii)} Rare fluctuations in a time series of positive couplings extending from the core to the cloud trigger the quakes.
The physical picture emerging highlights
the similarity of evolution and physical aging of complex materials. The ubiquitous role  of hierarchies in complex 
 dynamics~\cite{Simon62,Sibani13a} suggests that similar conclusions might hold
 beyond the TNM.\\
 
\noindent {\bf Background.}
Our  results  are based on simulations  performed at the SDU horseshoe cluster, using  C code  developed 
from scratch. Detailed  information on the model parameters, the initial conditions, and how
to generate the couplings can be found  in {Ref.~\cite{Anderson05}}, which should be consulted for further details. 
For convenience, some definitions and known properties are given below.

The TNM's  variables are binary strings of length $K$, i.e. 
points of the $K$ dimensional hypercube. Variously called \emph{species} or \emph{sites}, 
these are populated by  \emph{agents} or \emph{individuals}, which  reproduce
 asexually in  a way occasionally affected by random mutations. Only a tiny fraction of the possible species 
 ever becomes populated during simulations lasting up to one million generations. The  \emph{extant species}, 
 i.e. those with non-zero populations at a given time, are collectively referred to as \emph{ecosystem},
and their number as \emph{diversity}. 
With  probability $\theta$,
a  pair  $(a,b)$ of species has  non-zero couplings, ($J_{ab}$, $J_{ba}$),  describing how $b$ affects the reproductive  ability of $a$ and vice-versa.
{Empirically, the distribution of the generated couplings is well described 
by the  Laplace double exponential density  
$
p(x) = \frac{1}{2a} e^{-\lvert x - \overline{x} \rvert/a}.
$}
The parameters $\overline{x}$ and $a$ are  estimated to $-0.0019$ and $0.0111$, respectively.
   Extant species cluster 
together, their  closeness expressed by the \textit{Hamming distance}, the number of bits by which their  strings differ.

 Let $\mathcal{S}$,  $N_b(t)$ and $N$
 denote the ecosystem,  the population size of species $b$, and 
 the total population   {$N(t) = \sum_b  N_b(t)$}.
  An individual of type $a$ is chosen  for reproduction with probability $n_a = N_a/N$, and 
  succeeds   with probability %
$p_{\rm off}(a) = 1/(1 + e^{-H_a})$, where %
\begin{equation}
 H_a(t) = -\mu N(t)  + \sum_b j_{ab}(t),
\label{eq:Pfunc}
\end{equation}%
and where
\begin{equation}
j_{ab} =\frac{N_b}{N} J_{ab}= J_{ab} n_b
\label{dens_w_c}
\end{equation}%
is a density weighted coupling.
In Eq.~\eqref{eq:Pfunc}, $\mu$ is a positive constant. Letting $p_{\rm mut}$
 be  the mutation probability  per bit, parent and offspring differ by $k$ bits
 with  probability   Bin$(k;K,p_{\rm mut})$, the binomial distribution. Death occurs with  probability $p_{\rm kill}$
and  time is given in \emph{generations}, each equal to  the number of updates needed for all  extant individuals to
die.
Thus, with  population  $N$  at the end of the preceding generation,
the upcoming   generation comprises
 $N p_{\rm kill}$  updates. 
 The parameters used are always  $K=20$, $\mu=0.10$, $\theta =0.25$, $p_{\rm kill}=0.20$,  $p_{\rm mut}=0.01$,
and the initial condition invariably consists of  a single  species populated  with 500 individuals.

\begin{figure}[htb]
\centering
\includegraphics[width=.49\textwidth]{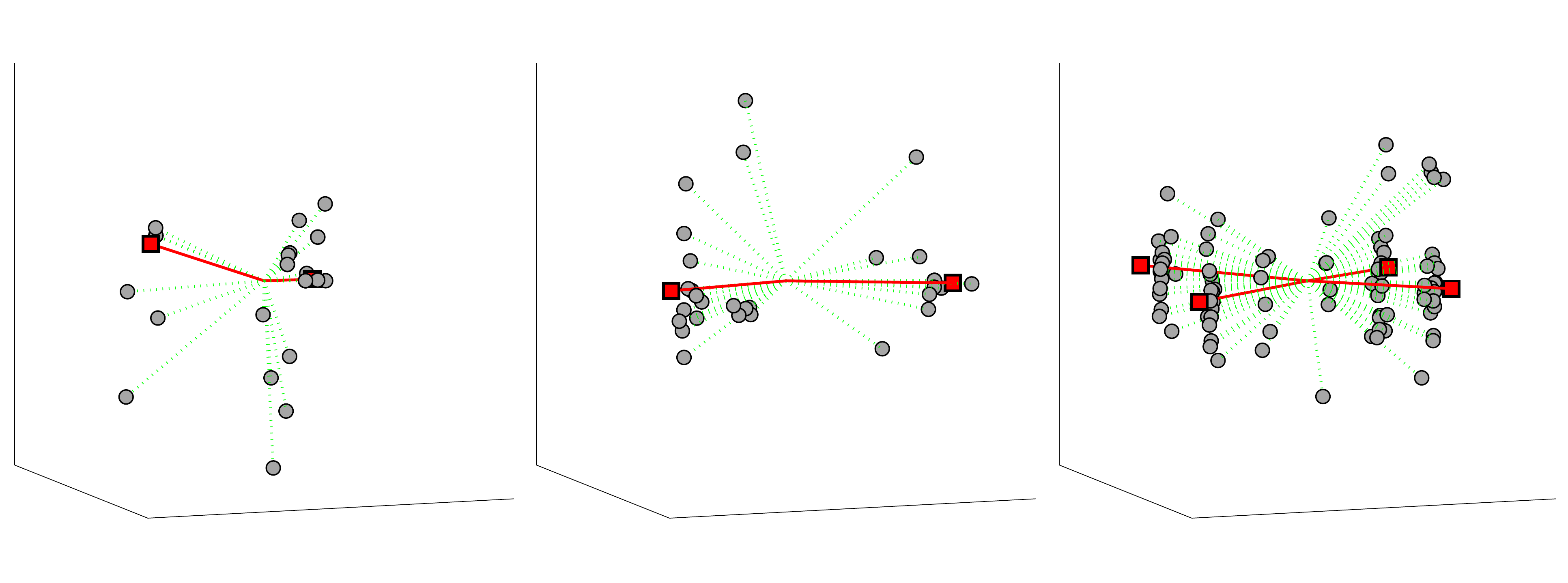}
\vspace{-3.5mm}
\caption{Core (red squares ) and cloud (gray circles) species at different system ages. 
All graphs drawn on the same scale.  {The red lines are  guides to the eye, showing the growing separation between core species.}}
\label{fig:3d_ecosys}
\vspace{-3.5mm}
\end{figure}

\noindent {\bf Core and cloud.}
\emph{Core species} have, by our definition, sizes exceeding  $5$\% of the most populous species.
  All together, they make up about 80\% of the  population. Other extant species, 
dubbed \emph{cloud species}, are sparsely populated, mainly by mutants of neighboring core species. 
A three dimensional visualization of the ecosystem is shown in Fig.~\ref{fig:3d_ecosys} 
{after $10^3, \; 10^5$ and $10^7$ generations}, with core and cloud species  marked by red squares and gray circles, respectively. Each core species is surrounded by its own cloud and both the number of core species and their distance, which reflects the Hamming distance, are seen to gradually increase as the system ages.

\begin{figure}[htb]
\vspace{-4mm}
\centering
\includegraphics[width=.45\textwidth]{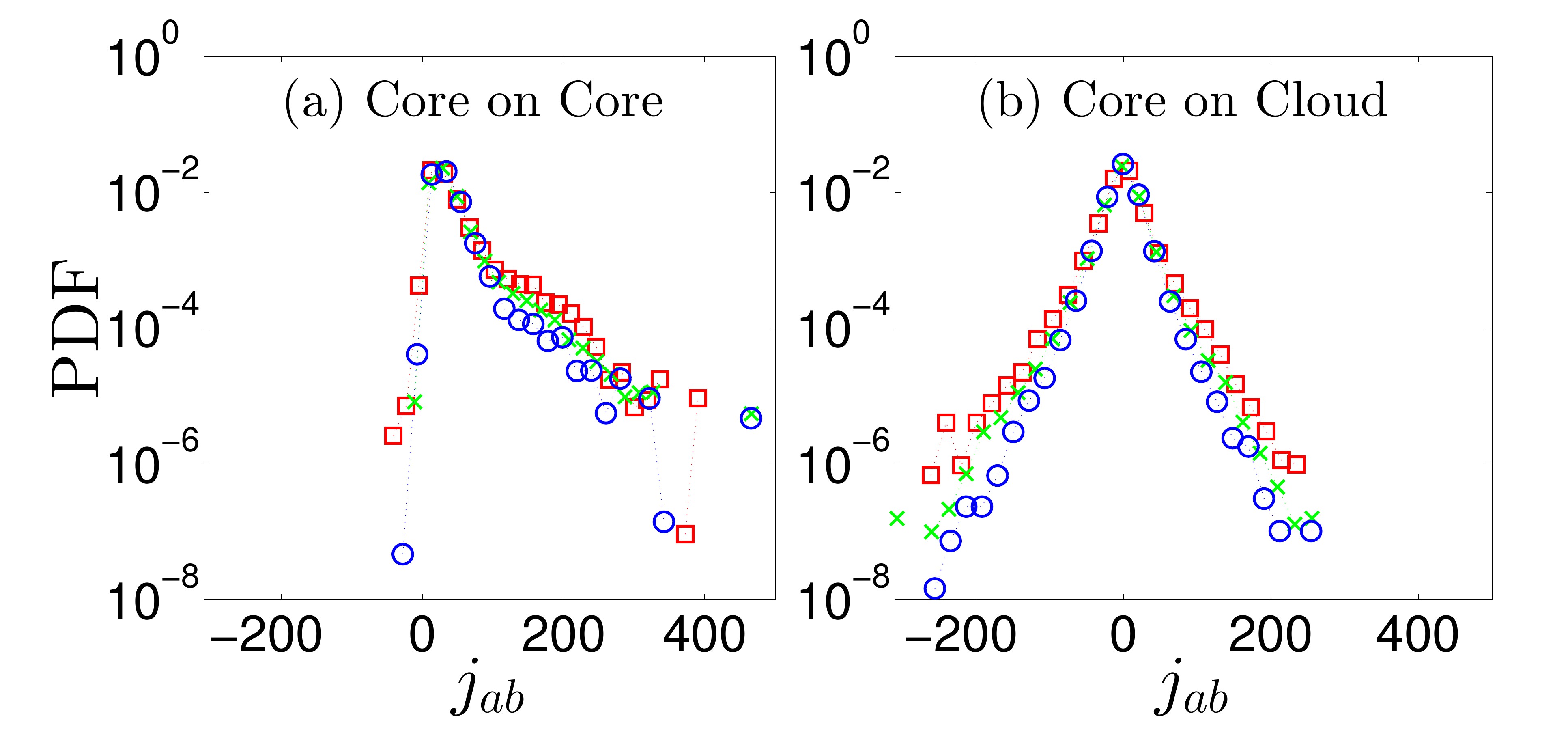}
\vspace{-2mm}
\caption{PDF of density weighted couplings: $t = 5\cdot 10^3$  (red squares), $t = 8\cdot 10^4$  (green crosses), $t = 10^6$  (blue circles).}
\label{fig:CoupStat}
\vspace{-3mm}
\end{figure}

Every panel of Fig.~\ref{fig:CoupStat} depicts the Probability Density Function (PDF) of the density weighted couplings, see Eq.~\eqref{dens_w_c},
 after $t = 5\cdot 10^3$, $t = 8\cdot 10^4$ and $t = 10^6$ generations.
The corresponding  data are sampled within qESS, where core and cloud are well defined. 
 Panel $(a)$ shows that negative couplings connecting core species are rare in a young system and then disappear. Hence, couplings do not specify
 trophic chains: A predator and  prey species stabilizing  each other
can   interact positively, while competing predators  can interact negatively.
  Couplings extending from core to cloud, panel $(b)$, feature a nearly symmetric PDF whose  width decreases with age. Core to cloud and cloud to cloud couplings
  have PDFs (not shown) similar to those of   arbitrary species.\\
  
\noindent {\bf Entropy, entropic barriers and  hierarchies.}
 In some thermalizing complex systems, increasing energy barriers
$b_n, \; n=0,1,\ldots$  separate the nested  \emph{metastable components}
 of a dynamical hierarchy,  see  e.g. \cite{Sibani89,Sibani93}.
When  starting out in  state $x_0$, surmounting the $n$'th barrier 
gives  access to a  component  ${\cal V}_n$  whose 
volume increases exponentially  with $n$.
To map out this  situation, the \emph{lid method}~\cite{Sibani99} introduces artificial and 
impenetrable energy barriers called `lids', which  allow the system to fully equilibrate  in 
the sub-volume of configuration space  below the lid.
As we argue, a  similar description holds  for the TNM,
 with   energy  barriers replaced by entropic ones. 

\begin{figure}[htb]
\vspace{-0.3cm}
\centering
\includegraphics[width=.45\textwidth]{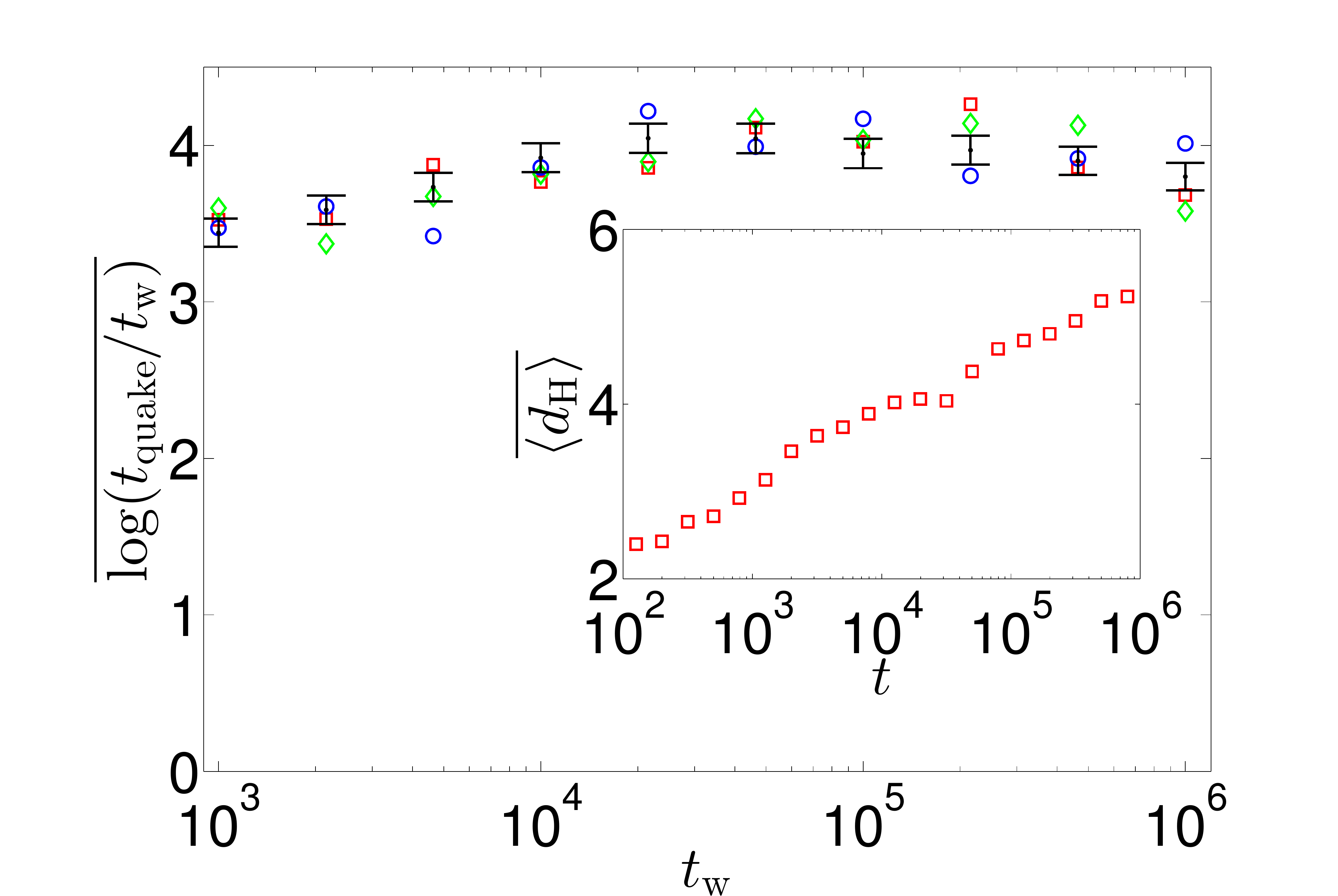}
\vspace{-2.5mm}
\caption{Main plot: The  data  with $1\sigma$ statistical error bars (black) show the  average of  $\log(t_{\rm quake}/t_{\rm w})$ vs. $t_{\rm w}$,
estimated using $2022$ trajectories. 
 The blue circles, green diamonds  and red squares  
are  based on different sub-samplings and 
illustrate  the statistical variation of the data. Insert:  Hamming distance from cloud species to the most populous core species, plotted on a   log scale
and averaged twice: over the cloud species and over $2022$ trajectories.}
\label{fig:cloud_size_av}
\vspace{-0.3cm}
\end{figure}

{The configuration volume  ${\cal V}$
associated to a qESS with 
$V$  extant cloud species and 
$N_{\rm cloud}$ individuals scattered among them is
approximately ${\cal V} \approx V^{N_{\rm cloud}}$. This 
 formula includes the (unlikely) case where  all cloud individuals
belong to the same species, which contradicts the definition of cloud species. Secondly,
 the core only serves to label the qESS and the
entropic contribution from its
(few) configurations is  neglected.
 The configurational  entropy is  } then
 $S = \log({\cal V}) \approx  N_{\rm cloud} \log(V)$, where $V$ 
   can be estimated via  the quantity $\overline{ \langle  d_{\rm H} \rangle}$, obtained by averaging the mean distance $\langle d_{\rm H}\rangle $ of cloud species to the most populous core species over the available ensemble of  $2022$ trajectories.
 To a good approximation,  the number of vertices at distance $k \ll K$ from a given vertex increases exponentially, 
 leading to $ V(t)  \approx \exp(\overline{ \langle  d_{\rm H} \rangle}/K)$. 
As shown in Fig.~\ref{fig:cloud_size_av}, $\overline{ \langle  d_{\rm H} \rangle} \propto  \log(t)$. Furthermore, we have 
checked  that $N_{\rm cloud} \propto \log(t)$.
 Hence, introducing  $t$ for the time scale of the qESS, we find
\begin{equation}
S(t)  \propto  \log(t)^2 \quad {\rm and }\quad  {\cal V}(t) \propto t^{a \log(t)},
\label{eq:entropydef}
\end{equation}%
where $a$ is a positive constant.
  As the  entropy increases and
   the free energy  correspondingly decreases  in time,  TNM  dynamics
 qualifies as  a spontaneous non-equilibrium physical process.
Importantly,     the source of  disorder lies
 entirely with  the cloud. 

 As discussed later, the fragility of  TNM ecosystems 
implies that a mutant  able to replicate successfully,  say mutant  $a$,
 quickly destabilizes the core. Consequently, a quake is triggered whenever 
$H_a \ge 0$, or equivalently, if $\sum_{b\in\mathcal{S}}j_{ab}(t) > \mu N(t)$. 
The  sum runs over all extant species but can safely be restricted to core species. 
In fact, as a mutant is most probably  connected to a single core species $c$, the criterion simplifies to%
\begin{equation}
j_{ac}(t) > \mu N(t).
\label{eq:lid}
\end{equation}%
Since $N$ on average increases, Eq.~\eqref{eq:lid} represents a rising bar for mutants to destabilize the existing core. 

\begin{figure}[htb]
\centering
\vspace{-4mm}
\includegraphics[width=.45\textwidth]{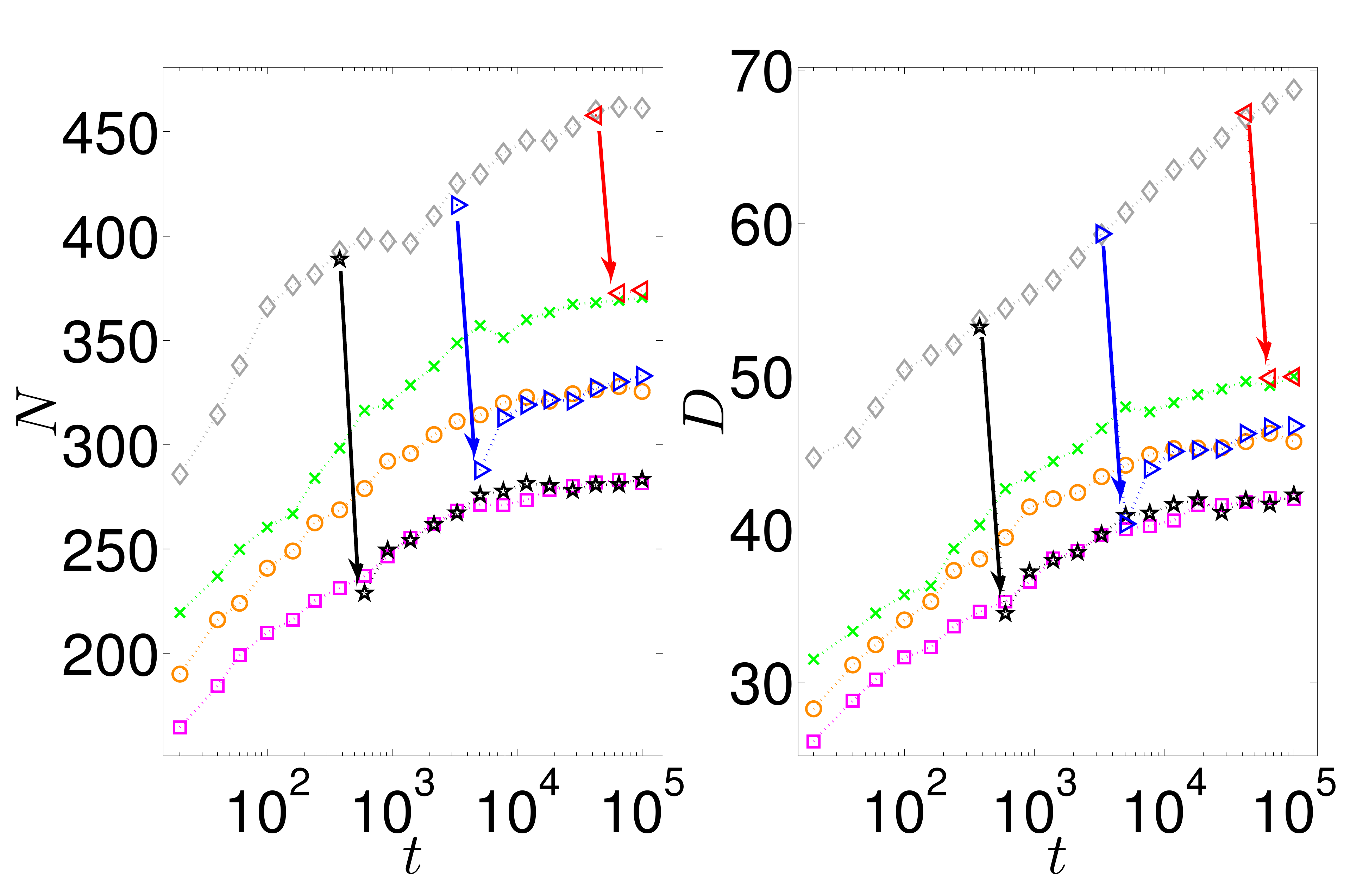}
\vspace{-3mm}
\caption{The two upper curves show: Left panel,   the average population, divided by two
for graphical reasons, right panel, the diversity, both plotted 
  vs. time. 
Lower  curves: same quantities with 
initially imposed lids.  $L=50,60$ and $70$ from  { bottom to top.}
Arrows show the fast decays which ensue 
when  the same lids are imposed  at times $t=5\cdot 10^2, 10^3$ and $10^4$.}
\label{fig:lideffekt}
\vspace{-3mm}
\end{figure}

As anticipated, we now modify the TNM by a `lid' rule: Each time a species $a$ is selected for reproduction, any coupling $J_{ab}$ entering $H_a$ and exceeding a preset value $L$ is set equal to $L$. Equation~\eqref{eq:lid} then entails $N(t)\le L / \mu$.  Figure~\ref{fig:lideffekt} shows that  our  lid rule  halts the evolutionary drift of the TNM, with the  stable levels of population and diversity reached depending on the size of the  lid  but not on the time of its deployment.  {Refs.~\cite{Rikvold03,Rikvold07,Rikvold07a,Murase10,Murase10a}, to which we 
shall return,  report and analyze   similar stationery fluctuation patterns.}
To estimate the level $N_L$ at which the population settles,  let $|{\mathcal{C}}|$ be the size of the core
and set  $J_{c'c} = L$, the highest value compatible
 with $J_{c'c} \le L$ in 
$
H_{c'} = \sum_{c = 1}^{|{\mathcal{C}}|-1} J_{c' c}\; n_c  -\mu  N_L \approx 0,
$
a slight over-estimate of the typical $H$ value of a stable core species $c$. This yields %
$
N_L \approx \frac{ L f_C}{\mu},
$
where $f_C=\sum_{c = 1}^{|{\mathcal{C}}|-1}n_c$. Based on data from Fig. \ref{fig:lideffekt}, Table \ref{tab:lidcalc} compares the measured population level $N_L$ with the estimate $L f_C/\mu$, showing that the latter overestimates $N_L$ by approx. 15\%. 
\begin{table}[!htb]
\vspace{-1mm}
\centering
\begin{tabular}{l l l l  r} 
\multicolumn{5}{c}{Lid on  the  interaction strengths}\\
\hline \hline
$t_{\rm lid}$ & $L$ & $f_C$ & $ N_L$ & $Lf_C/\mu$ \\ \hline 
     0 & 50 &  0.641 & 282 &  317 \\
     0 & 60 &  0.626 & 326 &  377 \\
     0 & 70 & 0.620 & 371 &  432 \\
   5$\cdot 10^2$ & 50 & 0.638 & 283 &  318 \\
  5$\cdot 10^3$ & 60 & 0.627 &  333 & 375 \\
 5$\cdot 10^4$ & 70 &  0.614 &  374 & 428 \\
\hline 
\end{tabular}
\caption{ Columns 1-3: $t_{\rm lid}$ is the time of lid deployment, $L$ is the lid value, $f_C$ is the
core population  fraction.
Columns 4-5: the population size $N_L$ for lid $L$, as obtained from Fig. \ref{fig:lideffekt}, and the corresponding estimate.}
\label{tab:lidcalc}
\vspace{-3mm}
\end{table}
To see why the lid locks the system size, note that the requirement for destabilization from species $a$, Eq. \eqref{eq:lid}, now reads $J_{ac} n_c > \mu N_L \approx L f_C$. 
Since  $L n_c > J_{ac} n_c$, destabilization requires $ n_c \ge  f_C$, which is impossible unless  the core  contains a single species.

According to  Eq.~\eqref{eq:entropydef}, 
 the configuration space volume available to a qESS at time $t$ 
is 
  ${\cal V}(t) \propto t^{a\log(t)}\propto t^{b N(t)}$, where $a$ and $b$ are positive constants. Furthermore,
 since the extant population $N$ grows linearly
with the lid, the configuration space volume   grows exponentially with it. This  already implies a hierarchical organization
of configuration space, with components mutually inaccessible on a time scale $t'$ or, alternatively, for a lid value $L'$,
merging  at  {  $t>t'$ or $L>L'$}. As a further check, we  reverse the process to see   
a component split: Consider  a  trajectory lasting 
$10^4$ generations. At  $t=10^3$ a lid $L=50$
is imposed and the system is allowed to relax  to a final state, 
labeled  by  its largest extant  species.
The  procedure is repeated $200$ times
with identical initial state and  random seed, but  resetting  
the seed to a different value each time  the lid is imposed. To improve the statistics, the whole process  is then  repeated for $94$ different
starting points. In approx. 75\% of the cases, more than 180 different end states are reached out of the 
$200$ possible. In the remaining 25\% of the cases, on average 65 different end states are reached.
In conclusion, the configuration space component available  after $10^3 $ generations contains a large 
number of sub-components with  different cores.\\

\noindent {\bf  Quake rate and qESS duration }
Since no macroscopic changes occur during  qESS, a coarse grained  description is naturally 
formulated  in terms of   Poissonian quake statistics~\cite{Sibani13} . Furthermore, based on the analysis of Ref.~\cite{Jones10},  the rate of quakes 
can be assumed to be $r_{\rm q}(t)=A/t$, where $0<A\le 1$ is a constant.
The condition $A<1$ excludes 
a partition of the system into statistically independent sub-systems, which is
fitting, as  all species are coupled  through  $N(t)$. 
 Let us finally assume that each quake leads to a random population change with average 
value $\mu_{\Delta}$. The average population after  $n$ quakes  is then $n \mu_{\Delta}$
which, averaged again  over  the probability that these  quakes occur in
the interval $(1,t)$,   finally yields %
\begin{equation}
N(t) =  \mu_{\Delta} A \log(t),
\end{equation}%
a logarithmic growth  in  qualitative agreement with our data.
An average over the population changes  incurred in all  quakes yields  $\mu_{\Delta}=105$. Using  the  logarithmic slope 
$\mu_{\Delta} A$ of the average population growth, fitted for  $t >1000 $, see Fig.~\ref{fig:lideffekt},
one finds $A=0.28$,  which is  close to the value $A=0.26$ obtained, as explained below,  from the
temporal  statistics of quakes. 
Assuming  a log-Poisson description for the latter,   the average number of quakes in $(t_{\rm w},t)$ is  $\mu_{\rm q}(t_{\rm w},t) = A \log(t/t_{\rm w})$, 
and the probability density for
 the first quake to happen at time $t>t_{\rm w}$ is $P_{\rm quake}(t_{\rm w},t) = A t_{\rm w}^{-1} (t /t_{\rm w})^{-A-1}$.
Averaging  $\log(t)$ over $P_{\rm quake}$  produces%
\begin{equation}
\overline{
\log  \left(
t/t_{\rm w}
\right)
}
 = \frac{1}{A} > 1.
\label{eq:logaverage}
\end{equation}%
The fair agreement with the estimate shown in  the main plot of Fig.~\ref{fig:cloud_size_av}
confirms that the  quake rate is  proportional to $1/t$. 
Note that the mathematical expectation 
of the qESS life-time $t-t_{\rm w}$  is undefined, and that the empirical average of the same quantity
correspondingly  features a huge scatter. 
For each trajectory, the entropic barrier $\Delta S (t_{\rm w})$ delimiting a qESS is the exponential of its duration.  Hence, Eq.~\eqref{eq:logaverage}
implies that 
the average entropic barrier grows  linearly with  $ \log(t_{\rm w})$.

The  main plot of Fig.~\ref{fig:cloud_size_av} is obtained 
by estimating  the time $t_{\rm quake} > t_{\rm w}$ at which a  core extant at time $t_{\rm w}$ is destroyed by 
a quake. To this end,   a large number of   mutants are generated and 
their ability to destabilize the core assessed. The procedure is carried out  for  ten  $t_{\rm w}$ values equidistant on a logarithmic scale 
stretching from $10^3$ to $10^6$ generations. The number of independent trajectories
 used for each age  varies between 907 (old systems) and 1663 (young systems). The variation 
 reflects  that,
 due to the finiteness of the system size,    some  ecosystems ---especially  old ones--- 
are  infinitely  stable, as   none of the mutants generated
can destabilize them.
Stable systems  cannot contribute to the quake statistics and are hence discarded.

 Mutant $a$ is deemed able to destabilize the core if $H_a > 5$, or equivalently, if its  reproduction probability exceeds
  $99.3\%$, a slightly more stringent requirement than the $H_a > 0$ implied by Eq.~\eqref{eq:lid}.
Glossing over the distinction between repeated and single mutations leading to  the same species, and using that core species by far are the main source of mutants, we let $M(l)$ be the total number of species at a distance $l$ from their common core ancestor $c$, and let $m_c(l)$ be  the number of those able to destabilize the core. Species $c$ is chosen for reproduction with probability $n_c$ and succeeds with probability $p_{\rm off}(H_c)$. The probability of producing a mutant at a distance $l$ is $P(X_{\rm mut} = l) = {\rm Bin}(l;K,p_{\rm mut})$ and the likelihood of hitting a destabilizing species under mutation of $c$ is $m_c(l)/M(l)$. All the above events being independent, the probability of destabilization per reproduction step at age $t_{\rm w}$ is%
\begin{equation}
p_{\rm quake}(t_{\rm w}) = \sum_{c \in \mathcal{C}} n_c \; p_{\rm off}(H_c) \sum_{l} P(X_{\rm mut}=l) \frac{m_c(l)}{M(l)}.
\end{equation}
The outer sum is over all core species and the inner one over all possible distances between mutant and parent species. 
Destabilization in $t$ attempts follows the geometric distribution with PDF $p_{\rm quake}(t_{\rm w}) (1-p_{\rm quake}(t_{\rm w}) )^{t-1}$, 
and the average number of attempts required is thus $1/p_{\rm quake}(t_{\rm w}) $. This leads to the estimate%
\begin{equation}
t_{\rm quake}(t_{\rm w}) =  t_{\rm w} + \frac{1}{p_{\rm quake}(t_{\rm w})}
\label{eq:lifetime}
\end{equation}%
for the  time at which a  core extant at time $t_{\rm w}$ is destroyed by a quake. 
Averaging over 2022 independent trajectories  the  logarithm of
$t_{\rm quake}/t_{\rm w}$   yields the main plot in Fig. \ref{fig:cloud_size_av}. 
The estimate   $A = 0.260\pm0.002$ is obtained straightforwardly from the latter.\\

\noindent {\bf Quake triggering fluctuations.}
After establishing  that a limit on   the range of $H_a$ stops the evolution of the TNM, we detail
 how   quakes are  triggered. To this end, data are needed with a temporal
resolution $200$ times higher than  in the rest of this work. These  data are collected
  for short intervals straddling   the approximate position of the quakes. The procedure is repeated
 for 100 trajectories lasting  up to $10^5$ generations and comprising at least $4$ quakes.

\begin{figure}[htb]
\vspace{-4mm}
\centering
\includegraphics[width=.45\textwidth]{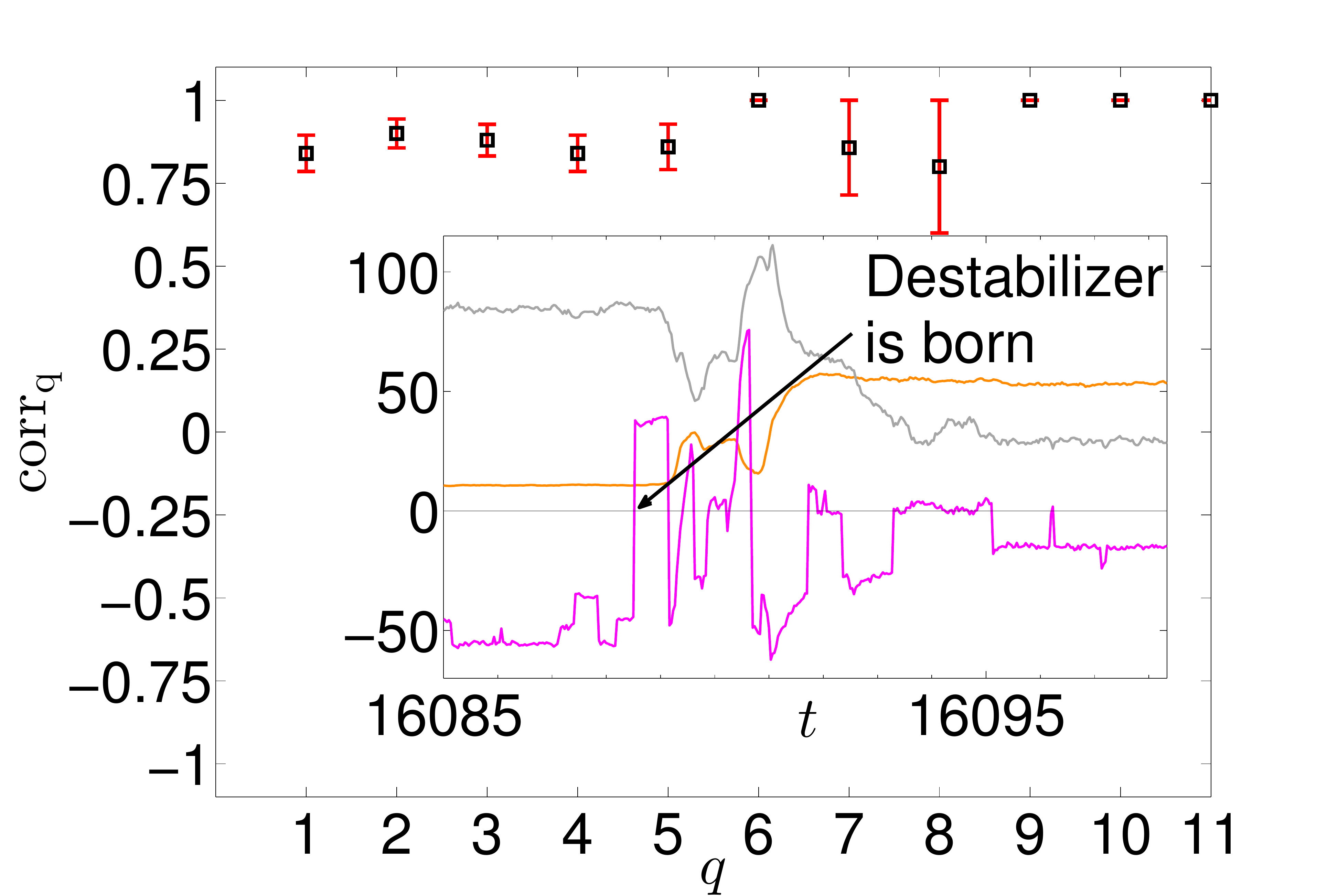}
\vspace{-3mm}
\caption{Main plot: Correlation between the  first zero crossing  of the 
trigger function  $T(t)$ to   positive values
  and a subsequent large change in  population and/or Center-of-Mass (COM).
   Insert: $T(t)$ vs. time, showing its first zero crossing and the concomitant
  birth of  a  destabilizer species at $t = 16,096$.
  Also included are  traces  of population (gray) and COM (orange) scaled down for  convenience.}
\label{fig:highRes_analysis_w_insert}
\vspace{-3mm}
\end{figure}

\begin{figure}[htb]
\vspace{-5mm}
\centering
\includegraphics[width=.45\textwidth]{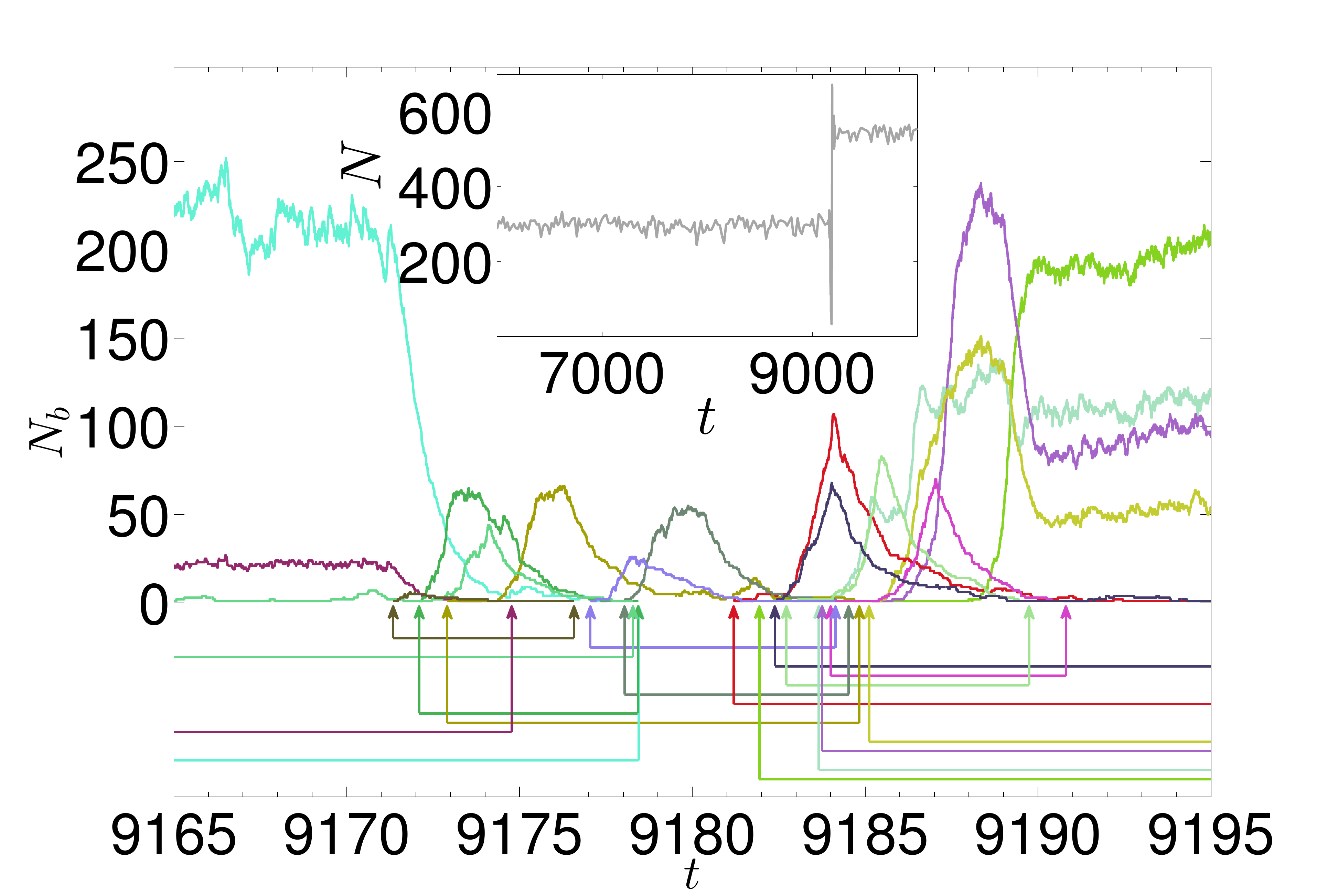}
\vspace{-3mm}
\caption{Main plot: The noisy curves show  population  vs. time for  
selected  species able to reproduce  during  a quake.    Their  birth and extinction 
 are marked by color coded vertical arrows connected by horizontal lines. 
The species appearing at $t_q \approx 9171$ (dark green curve) marks the beginning of the end. 
Insert:  the total population vs. time  before and after the quake.}
\vspace{-0.3cm}
\label{fig:individual_pop_devel}
\end{figure}

To gauge the highest contribution to the $H$ values of the cloud species, we define  the time dependent `trigger' function%
\begin{equation}
T(t) = \max_{a \in {\rm cloud}}
\left\{
\sum_{c \in {\rm. core}}
J_{ac} n_c
\right\}
-\mu N(t),
\end{equation}%
 depicted as the magenta curve in the inner plot of Fig.~\ref{fig:highRes_analysis_w_insert}. Initially fluctuating
 well below zero, $T$ suddenly jumps above zero at $t=16,096$. Soon thereafter, the population  
 (the gray curve shows the population scaled down for convenience), decreases dramatically,  
 which is a sign of destabilization. A further indication   is the clearly visible change in the system's Center of Mass,
 (orange curve, also scaled down), the latter obtained by mapping the bit string of each species into a real number
 and interpreting the corresponding population as a mass.
The quake extends from $t=16,096$ to $t=16,119$, 
a minuscule   interval during which $T(t)$ oscillates erratically before returning to  the negative values characterizing 
the new qESS. The situation just described pertains to a single trajectory. To ascertain its general validity, we define a Boolean matrix $B_{i,q}$, where $i$ and $q$
 index trajectories and  quakes within a trajectory, respectively. 
 For each  $i$, $B_{i,q} = 1$ if the $q$'th quake is preceded by a positive fluctuation of $T(t)$ and $-1$ if not. 
Averaging over all  trajectories yields the function ${\rm Corr}(q)$ depicted in the main plot of Fig.~\ref{fig:highRes_analysis_w_insert}. 
${\rm Corr}(q) = 1$ 
would indicate a perfect correlation.
The values shown  are a  bit lower since 
 in rare cases,  values of $T(t)$ lingering slightly below
zero can  trigger quakes.
Figure~\ref{fig:individual_pop_devel} further details how the sudden growth of one mutant species induces a quake. 
Before $t_q \approx 9171$ two stable core species are present. 
The arrival of a new species (dark green curve) at $t_q \approx 9171$ starts a quake which ends  in   a new stable configuration 
 at $t \approx 9190$. The latter  has twice as many  core species as the old one and almost twice the population. 
During the quake many  mutants  gain population and   disappear. The populations of  a  few of these mutant species are plotted  in the figure vs.
time, with arrows pointing to the instants at which they appear and again disappear.
Quakes are generally short and turbulent periods during which old species disappear and new ones gain 
foothold. In the present example a species which never had more than 10 individuals manages to destroy a core stable 
through many generations. The surprising fragility of the TNM ecosystems appears similar to the fragility 
of real ecosystems with respect to the introduction of new \emph{invasive species}.\\

\noindent {\bf Discussion and outlook.}

 Focusing  on    the  non-stationarity nature of  the  dynamics,
 we have  placed the TNM
in   a  wide class  of hierarchically organized~\cite{Simon62}  metastable systems, 
alongside with aging glassy materials~\cite{Sibani13a}.
During a   qESS the dynamics is stationary   within a 
  component of the  hierarchy and
  full stability is achieved by imposing
a  `lid'   curtailing 
the Laplace distribution of the couplings into one  supported in a finite interval. 
 {A  coupling distribution with finite support, as used in 
Refs.~\cite{Christensen02,Rikvold03,Rikvold07,Rikvold07a,Murase10,Murase10a},
seems therefore 
 crucial for  reaching  a  stationary regime within accessible time scales.
 This  
regime   is studied in detail and given a biological interpretation 
in~\cite{Rikvold03,Rikvold07,Rikvold07a,Murase10,Murase10a}.
An interesting open question concerns the dynamical effect of  a  
broad, e.g. power-law,  distribution of couplings,  }
 
Hierarchical systems    admit a coarse-grained dynamical description in
terms of quakes  connecting   metastable states.
These quakes constitute   a Poisson process~\cite{Sibani13},  whose  average depends on a difference of 
`stretched  times', with the stretching function being  a  logarithm in the TNM case. 
Such  `log-Poisson' processes   arise 
when   quakes are triggered by record breaking fluctuations~\cite{Sibani13}
 but can    also  follow  from the  gradual increase of  dynamical barriers
 in a hierarchy~\cite{Sibani13a}.  
 
 Can  `fitness'  be an emerging property of the TNM? At the individual level the answer is
  negative by construction. At the systemic level, e.g. the individual level of a coarser description, 
the  only measure of  success is long-term core stability. The latter can   result from all mutants receiving 
negative interactions  and  hence being unable to 
reproduce.  Conversely, 
depending on whether a core or a cloud species is at the receiving end, positive interactions
 underlie the stability of the  core or cause its eventual demise. Since
the interactions linking  core  species are nearly irrelevant for stability, 
evolutionary success  is not a function of   the state of the core.
Hence,  coarse graining the latter   into a `compound'  species does not lead  to a fitness based 
evolution model similar to    e.g. the Kauffman's ~NKC  model \cite{Kauffman91}.
{The result more closely resembles a neutral model of evolution, see \cite{Duret08}
and references therein,
with the proviso that the rate of genetic drift is in our case decelerating 
if the environment stays  constant.}

 In our log-Poisson description, a qESS  has infinite expected life-time, 
while the expectation value  of the logarithm  of its duration depends on 
 how long  the core has existed. This weak predictive ability is reminiscent of, and might even supply 
a formal mathematical basis for,  the \emph{ontic openness}~\cite{Jorgensen12} of real ecosystems.

The TNM population size depends on $\mu$ and even though a $\mu$-cycle 
(of moderate amplitude) will eventually  restore the population at its original level, 
we expect that an increase followed by a decrease will modify the core, while the
 inverse process will leave it unchanged. In other words, the first process lets the 
 system explore  new parts of its hierarchical 
configuration  space, similarly to rejuvenation\cite{Berthier02,Sibani04}, while the
 second keeps  a memory of the past state. Possibly, partial randomization achieved by 
 a periodic variation of the environment can accelerate the pace of evolution and, in the
  context of the extended TNM with  spatial features~\cite{Lawson06}, lead to the formation of new structures at a higher level of aggregation.
Finally, since  the bit strings of the TNM can code for strategies of economic agents~\cite{Robalino12},
our analysis might  be relevant  for understanding the optimal balance between continuity and innovation  in  human societies.

\noindent {\bf Acknowledgements}.  
P.S. is grateful to Henrik  Jeldtoft Jensen and Per Lyngs Hansen  for  many inspiring discussions
 {and thanks  two anonymous referees for their constructive advice}.


\end{document}